# Integrated STEM in Elementary Grades Using Distributed Agent-based Computation

Pratim Sengupta, Gokul Krishnan and Mason Wright
*Mind, Matter & Media Lab, Vanderbilt University – Peabody College, Nashville, TN, USA*
{pratim.sengupta, gokul.krishnan, mason.wright}@vanderbilt.edu

Abstract: We investigate how the integration of visual agent-based programming and computationally augmented physical structures can support curricular integration across STEM domains for elementary grade students. We introduce ViMAP-Tangible, a socio-technically distributed computational learning environment, which integrates ultrasonic sensors with the ViMAP visual programming language using a distributed computation infrastructure. In this paper, we report a study in which $3^{rd}$ and $4^{th}$ grade students used ViMAP-Tangible to engage in collaborative design-based activities in order to invent "drawing machines" for generating geometric shapes. The curricular activities integrate engineering practices such as user-centered design, mathematical reasoning about multiplication, rates and fractions, and physical science concepts central to learning Newtonian mechanics. We identify the key affordances of the learning environment and our pedagogical approach in terms of the relationship between the structural elements of students' physical constructions and computational models, and their learning outcomes, both in terms of computational thinking, and the domain-specific, mathematical and scientific knowledge that they began developing.

## 1. INTRODUCTION

Integration of the individual domains of science, technology, engineering, and mathematics (STEM) is now recognized as a central pedagogical aim of engineering and science education reform at the K-12 level (Nathan, Srisurichan, Walkington, Wolfgram, Williams, & Alibali, 2013; Berland, 2013). STEM integration is considered in the US education policy statements to be necessary for several objectives: a) supporting STEM education, including the preparation of future STEM researchers; b) for developing informed citizens; and, c) for supporting workforce development in an increasingly complex economy (Katehi *et al*., 2009; NRC, 2007, 2010; Nathan *et al*., 2013).

Integrated STEM necessitates integrating diverse domains by highlighting *big ideas* that transcend these different domains (Nathan *et al.*, 2013; Schunn, 2009; Roehrig, Moore, Wang, & Park, 2012). However, the generation of these ideas involves material agency, conceptual agency and participation in a community of practice. For example, engineering educators have argued that some of the most essential "skills" in engineering "arise out of engagements not only with formal representations, but also with tools, materials, and other people" (Johri & Olds, 2011, p. 163). Similarly, historians, philosophers and sociologists of science have shown that the development of scientific knowledge (e.g., *big ideas* such as laws of physics) is deeply intertwined with the invention of representational systems and tools, as well as the development of communicative representational practices (e.g., modeling) (Giere, 1999; Pickering, 1993). The representational systems and tools include both semiotic systems (e.g., calculus and computational modeling languages) and mechanical devices (e.g., bubble chambers and particle accelerators). This is known as the *Science as Practice* perspective, and has been adopted as a key pedagogical framework for K-12 science education in the US (NRC, 2008).

Given this background, we believe that the focus on Integrated STEM is synergistic with the recent focus on computational thinking (Wing, 2006, 2010; NRC, 2010). Wing (2006) described computational thinking as a general, analytic approach to problem solving, designing systems, and understanding human behaviors. Sengupta, Kinnebrew, Basu, Biswas & Clark (2013) argued that computational *thinking* is evident in the form of *epistemic and representational practices* such as problem solving, design, programming, and modeling.

Pedagogy that supports the development of representational practices associated with computational thinking can bring together different domains in science, such as biology and physics, in middle school classrooms through the use of agent-based, visual programming languages designed specifically for modeling scientific phenomena (Sengupta *et al.*, 2013). In agent-based programming, a user creates a computer program by using simple rules to command the movement and behavior of computational agents, e.g., the Logo turtle (Papert, 1980; diSessa, Abelson & Polger, 1991; Repenning, 1993; Kelleher & Pausch, 2005; Resnick *et al.*, 2009; Sengupta & Farris, 2012; Sengupta *et al.*, 2013). In this paper, we extend this argument and propose that a particular form of agent-based programming and modeling, in which control of a computational agent is socio-technically distributed, can be leveraged to integrate diverse STEM domains for children in elementary grades.

The learning environment we present here can be best understood as a *socio-technically distributed* activity system. This is because ViMAP-Tangible distributes the control of a single computational agent *socially* between two collaborating users, and *technologically*, between two physical machines and a virtual algorithm. The goal of each student dyad in this study was to design, build, test and refine a hybrid (computational and physical) computational machine for generating geometric shapes. Our pedagogical approach emphasized User-Centered Design (UCD), i.e., students were asked to design their machines with *usability* (Norman, 1998) as a key focus, which has been shown to be a crucial element of product engineering but challenging to implement pedagogically.

Our paper makes three contributions. First, we present a pedagogical framework for integrating key engineering practices – UCD, collaboration and computational thinking - with math and science education. To this end, we present a theoretical framework that integrates relevant literature from multiple domains: computational thinking, Integrated STEM, Agent-based and Tangible Computation in education, User Centered Design, Design-based Learning and Collaboration in engineering practice and education. Second, we present a technological innovation in the form of ViMAP-Tangible, and a set of curricular activities, which were designed to bring about such integration. Third, while recent efforts have focused on designing and implementing Integrated STEM curricula at the college level (Sanders, 2009), middle school (Berland, 2013) and high schools (Nathan *et al.*, 2013), we demonstrate that younger children ($3^{rd}$ and $4^{th}$ graders) can be brought into the fold of Integrated STEM education that also includes a focus on developing computational thinking.

## 2. THEORETICAL FRAMEWORK

## 2.1 Computational Thinking in K-12

Computational thinking is an increasingly ubiquitous epistemic and representational practice in all fields of science and engineering (Wing, 2006; NRC, 2010). As Sengupta *et al.* (2013) pointed out, computational thinking draws on concepts that are fundamental to computing and computer science, but also includes practices (e.g., modeling, abstraction, reformulation, simulation, verification) that are central to a large number of scientific, engineering, and mathematical disciplines. This sentiment is also reflected in the model ACM K-12 computer science curricula for middle schools (Tucker *et al.*, 2003), and the recently concluded National Academy of Education panel on computational thinking (NRC, 2010), which argued for integrating computational thinking with existing K-12 curricula in other domains such as mathematics and science.

Wing (2006) argued that a key characteristic of computational thinking is design-based thinking. Design is a form of problem solving in which thinking, tool manipulation, and materials are reflected in the iterative construction of an artifact (Bucciarelli, 1994; Simon, 1969; Perkins, 1986). From a pedagogical perspective, design challenges provide learners opportunities for testing and revising their developing conceptions and understanding, and interweave action and development with reflection and refinement to facilitate deep learning (Kolodner *et al.*, 2003). Researchers have shown that students' construction failures, when scaffolded appropriately, provide additional opportunities for learning (Kolodner *et al.*, 2003; Papert, 1980; Harel, 1990; Penner, Lehrer & Schauble, 1998).

## 2.2 Integrated STEM As Pedagogy

Integrated STEM has been defined as technological and engineering design-based learning approaches that intentionally integrate concepts and practices of science and/or mathematics education with content and process of technology and/or engineering education (Sanders, 2009; McCulloch & Ernst, 2013).

As Nathan *et al.* (2013) pointed out, integration necessitates a pedagogical approach in which fields are *integrated*, rather than merely combined (cf. Dyer, Reed, & Berry, 2006; Satchwell & Loepp, 2002). The emphasis on "integration" implies that diverse fields of knowledge and practice should be merged in a manner that reveals big ideas that transcend specific disciplines (Nathan *et al.*, 2013; Schunn, 2009; Roehrig, Moore, Wang, & Park, 2012).

Nathan *et al.* (2013) argued that big ideas that are regarded as invariants in math and science are represented using different inscriptions, both material and semiotic, as well as in different social interactions, such as in lectures or group work. As such, Nathan *et al.* (2013) argued that STEM integration in a high school level project-based engineering classroom can be viewed as the production and maintenance of *cohesion* of invariant relations across the broad range of representations that exist in the engineering classroom.

Similarly, Berland also argued that STEM integration can be brought about by a particular class of activities that she termed *STEM-design challenges* (Berland, 2013). In such activities, students are posed an engineering design challenge that can only be completed when relevant math and science concepts are applied (e.g., Coyle, Jamieson & Sommers, 1997; Fortus *et al.*, 2004; Kanter, 2010; Kolodner *et al.*, 2003). In such activities, pedagogically, these concepts represent domain-specific learning goals in science and math.

One of our central goals in this paper is to demonstrate that computational representational practices that are supported by agent-based programming and the design and development of usable physical control mechanisms for controlling agent-behaviors, can integrate representational practices and conceptual development across multiple STEM domains.

## 2.3 Agent-Based Visual Programming & Tangible Computation for Children

The literature on designing agent-based programming languages and environments for novice programmers highlights the following affordances. First, agent-based programming has been shown to be intuitive for novice programmers, as it leverages learners' embodied intuitions about movement in space (Papert, 1980). Second, agent-based programming can also help children learn scientific concepts in physics and biology (diSessa, Abelson & Ploger, 1993; Repenning, 1993; Sengupta & Farris, 2012; Sengupta *et al.*, 2013). Third, agent-based programming can help develop computational literacy through the design of self-expressive digital narratives and games using visual agent-based environments such as Scratch (Maloney *et al.*, 2004) and Alice (Conway, 1997).

Our choice of visual programming as the *mode* of programming is grounded in the literature on computer science education. Children find it difficult to understand the syntax and semantics of programming (Spohrer & Soloway, 1986; Perkins, 1988). They also find it challenging to effectively control the flow of a program using loops and conditionals (du Boulay, 1989). Researchers have also found that alleviating syntax problems helps students focus on the semantic ones (Hohmann, 1992; Soloway, 1993; Anderson, 1989; Mannila, Peltomaki & Salakoski, 2006). Visual programming – in which students construct programs using graphical objects in a drag-and-drop interface– has been shown to be effective in alleviating these difficulties (Kelleher & Pausch, 2005). Examples of agent-based visual programming environments are AgentSheets (Repenning, 1993), StarLogo TNG (Klopfer, Yoon, & Um, 2005), Scratch (Maloney *et al.*, 2004), ViMAP (Sengupta, Farris & Wright, 2012) and Alice (Conway, 1997).

Researchers have also started focusing in on the integration of tangible computation with agent-based modeling and programming for novice learners (Suzuki & Kato, 1995; Blackwell, 2003; Horn & Jacob, 2007; Blikstein & Wilensky, 2009). Blikstein and Wilensky (2009) showed that linking multi-agent computer models with real-world phenomena by using sensors could enable undergraduate students to learn authentic scientific and engineering concepts and practices. Tangible programming languages such as AlgoBlock (Suzuki & Kato, 1995) and Tern (Horn & Jacob, 2007; Horn *et al.*, 2011) have also been used to teach young children programming.

In AlgoBlock, lexical elements of the Logo language are assigned to sealed metal boxes, about 20 cm on each side. These blocks can be assembled by plugging each block, via connectors on the sides, into neighboring blocks. These blocks are connected to the computer using a wired interface. A complete Logo algorithm could be constructed by assembling a sufficient number of blocks on a tabletop, and the results of running the program are visible in the form of an animated submarine on screen (Suzuki & Kato, 1995). Horn and colleagues developed Tern, in which users construct programs by arranging and organizing wooden blocks with computer-vision fiducials (black and white symbols), which are then scanned by a program to generate a Logo algorithm. Tern has been used effectively to teach young learners programming and robotics, both in informal and formal settings (Horn, Crouser & Bers, 2011; Horn & Jacob, 2007).

While tangible and visual programming offer two different interactional modes of programming for the learner, Horn, Crouser & Bers (2011) have argued for introducing a hybrid approach in which learners (users) can elect to work either using tangible blocks, or by using a visual (graphical) programming interface, to generate the same algorithm. They found that such a hybrid approach was more advantageous than using either approach individually.

## 2.4 User Centered Design: Practice and Pedagogy

User Centered Design (Norman, 1998; Norman & Draper, 1986) emphasizes the importance of understanding the needs of the users in order to design usable systems. Norman (1998) argued that central principles of designing for people to support understandability and usability are: a) providing a good conceptual model to the user, which will allow them to predict the effects of their actions on the designed system; and b) making things visible, i.e., by carefully considering the relationship between the design of "controls" of a system and its "placement" (location) so that it makes the function of the control intuitively available for the user.

However, there is variability in how design researchers have defined and operationalized UCD: while there is a general agreement that UCD requires paying attention to the needs of the user, and involving the user in the system design process, there is relatively less agreement on how user involvement can be accomplished (Vredenburg, Mao, Smith, & Carey, 2002; Gulliksen, Göransson, Boivie, Blomkvist, Persson, & Cajander, 2003; Karat, 1996). Vredenburg et al. (2002) defined UCD as the practice of the following principles: the active involvement of users for a clear understanding of user and task requirements, iterative design and evaluation, and a multi-disciplinary approach. Gulliksen et al. (2003) further elaborated on the principles and argued that in addition to involving the user throughout the design process, the following principles are necessary to be enacted in practice in order to support UCD: 1) rapid prototyping during the early phases of the design process; 2) a cyclic iterative process of designing solutions interwoven with evaluation; 3) multi-disciplinary design teams that bring together distributed expertise for the various design components; and 4) an integrated design process, in which the system, the work practices, on-line help, training, organization, etc. should be developed in parallel.

In this paper, we adopt Gulliksen et al.'s (2003) definition of UCD as a key element of our pedagogical approach. There is some evidence that designing for instructional use can act as a productive pedagogical model for K-12 science education (Carver et al., 1992; Harel, 1990; Brown and Campione, 1993). Brown and Campione (1993) noted that fifth and sixth graders developed a deeper understanding of science concepts while they sought causal explanations to incorporate into HyperCard documents they developed to teach their classmates. A few studies have also focused on children designing agent-based instructional software for mathematics (e.g., Harel, 1990) and instructional games in science and mathematics (e.g., Kafai et al., 1998). In all of these studies, students not only developed a deeper understanding of the target science or math concepts, but also developed substantial expertise in programming.

An interesting finding across these studies is that children find the consequentiality of their design projects in terms of designing for use to be quite motivating, but at the same time, they do not regard the involvement of users as a useful component of their design process (Carver et al., 1992; Kafai et al., 1998). Carver et al. found that "getting someone to try out the presentation" was regarded by middle school students as one of the least important tasks to accomplish during their design process; instead, they believed that the designers themselves could act as users during the design process. Similarly, in Kafai et al.'s work, children who designed educational software did so largely without involving real users in their design process. This in turn resulted in the design of user interfaces that were confusing for the real users (Kafai et al., 1998). This is in striking contrast to Norman's (1998) famous dictum that because the interface guides the interactions between the user and the product, it should therefore guide the design of the rest of the product.

These studies suggest that the involvement of users during the design process therefore requires explicit instructional scaffolding by the teacher. In fact, once "real" users tested the children's designs, Carver *et al.* (1992) found that the design documents designed by the children to scaffold user interaction with their designs increased greatly in terms of making explicit the connections between the different aspects of their design, as well as explaining how to use the designed artifact. In Norman's terms (Norman, 1998), one can therefore conclude that involvement of user feedback during the design cycle enhances the usability of the product, as it makes clear to the designers the need to *make things visible* to the user.

## 2.5 Collaboration in Engineering Practice & STEM Education

Collaboration and teamwork hold a significant place in engineering practice (Bucciarelli, 1994). As Johri (2012) pointed out, engineers increasingly work collaboratively around the globe; technology is a primary driver of such arrangements. Anderson, Courter, McGlamery, Nathans-Kelly and Nicometo (2010) studied engineering work across six engineering firms that examined engineers' self-valuation of important work characteristics. They found remarkable similarities across settings in which most engineers saw their work as collaborative problem solving, and greatly valued communication skills and working in a team.

In the field of education research, collaboration and cooperation have been defined as conceptually distinct processes. Roschelle & Teasley (1995) defined collaboration as "a coordinated, synchronous activity that is the result of a continued attempt to construct and maintain a shared conception of a problem" (Roschelle & Teasley, 1995; p. 70). Dillenbourg (1999) defined the distinction between collaboration and cooperation as follows: "In cooperation, partners split the work, solve sub-tasks individually and then assemble the partial results into the final output. In collaboration, partners do the work 'together.'" (p. 8).

Productive collaborative tasks create *positive interdependence* among learners, which can be understood as the coordinated activity of multiple people for accomplishing a specific learning objective (Kreijns, Kirschner & Jochems, 2003; Antle & Wise, 2013). White & Pea (2011) argued that collaborative tasks are likely to be most effective when they are sufficiently open-ended and complex to necessitate contributions from each member of a group (e.g., see Cohen, 1994), and when participants engage the task and one another in ways that sustain their diverse contributions (e.g., see Barron, 2003). In the literature in computer supported collaborative learning, collaboration is often fostered through variations on the "jigsaw" script in which each student only has access to part of the information (i.e. one piece of the puzzle) needed to solve a collaborative task (Miyake, Masukawa & Shirouzou, 2001).

Research suggests that collaboration can indeed enhance science and math learning through the creation of productive opportunities for shared inquiry and discourse. Pathak, Kim, Jacobson & Zhang (2010) showed that dyadic discourse in an ill-structured inquiry activity when exploring a scientific phenomenon using agent-based simulations, creates opportunities for reflective reasoning. Specifically in the context of learning kinematics using agent-based programming, researchers have showed that collaborative design of graphical, mathematical and computational representations of motion creates productive opportunities for within-group "conversations for conceptual change" (diSessa *et al.*, 1991; Sherin *et al.*, 1993). Similarly, in math education, researchers have focused on how learning in collaborative settings leads students to develop mathematical discourse by productively appropriating their group members' ways of talking or acting (Carlsen, 2010; Moschkovich, 2004; Radford, 2006; Lai & White, 2011).

Of particular relevance to our paper is the argument that meditational *tools* – that include both tangible and computational artifacts – can support collaborative learning by creating opportunities for each group member to attend to what the other is doing, by making actions visible and gaze observable in supporting collaborative meaning-making (Antle & Wise, 2013; Fernaeus & Tholander, 2006; Hornecker, 2005; Suzuki & Kato, 1995; Baker *et al.*, 1999; Suthers *et al.*, 2008). These studies showed that the presence of tangible artefacts in a shared *transaction space* (Hornecker, 2005) grounds the interaction between group members by providing a referential anchor for conversation, which can be referred to by using both verbal and gestural communication channels. Of direct relevance to our study, Fernaeus & Tholander (2006) found that when students worked collaboratively using a tangible programming language for learning agent-based programming, they formed subgroups dynamically, and furthermore, and these subgroups further collaborated with one another on the different activities to accomplish their goals.

# 3. THE VIMAP-TANGIBLE LEARNING ENVIRONMENT

## 3.1 The Representational Infrastructure

Three key elements of ViMAP-Tangible are:

(1) *Agent-based visual programming*: ViMAP (Sengupta, Farris & Wright, 2012) serves as the agent-based, visual programming language for our learning environment. ViMAP consists of library of graphical programming primitives designed specifically to support learning of mathematics, kinematics and biology, a construction zone where learners can generate an algorithm by using a drag-and-drop interface that is easy to understand for children, and a NetLogo environment (Wilensky, 1999) to allow learners to visualize the results of their algorithm. Similar to Alice and Scratch, ViMAP provides a drag-and-drop interface for constructing programs that is easy to use and understand for children. Commands in ViMAP include both domain-general abstractions (e.g., loops, conditionals), as well as domain-specific commands (e.g., for controlling speed, distance and acceleration of the turtles).

(2) *Tangible and gestural representation of digital information*: In ViMAP-Tangible, students can use gestures and/or mechanical devices in order to control different variables by linking them to sensors through their ViMAP programs. The ultrasonic distance sensors we used in this study measure the distance from the sensor itself to the nearest object in its field of operation. Some of the ViMAP programming primitives were designed specifically to allow users to set the values of different agent-variables (e.g., color, speed, acceleration, pen-width, horizontal and vertical displacements) based on the sensor-readings. For example, the "set <step-size> equal-to <sensor-reading>" command sets the distance the agent on the screen travels equal to the reading of the ultrasonic sensor. That is, when a learner uses this command in her/his algorithm, moving a hand towards the sensor (i.e., closer) will cause the turtle to travel a shorter distance.

(3) *Distributed control of virtual agent*: In ViMAP-Tangible, multiple students can simultaneously control the behavior of a single agent. In the version of ViMAP-Tangible we report here, control of the behavior and attributes of a single computational agent is distributed across two ultrasonic sensors, which are connected to ViMAP via an Arduino$^{TM}$ microcontroller (Figure 2). We believe that such a setting can foster *positive interdependence* (Antle & Wise, 2013), since it requires the coordinated action of both the group members in order to computationally implement and physically enact a successful and non-redundant control mechanism.

## 3.2 Curricular Activities

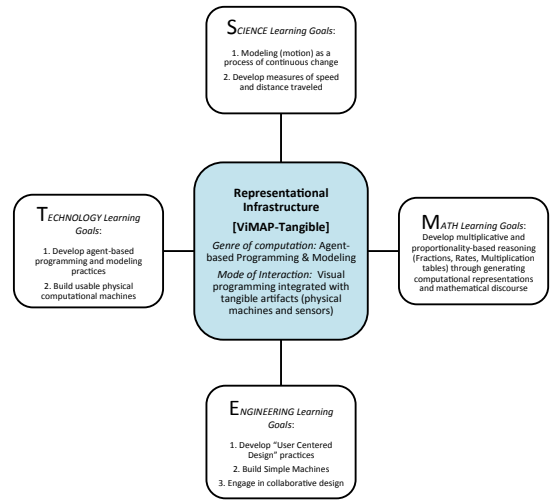

Figure 1: Domain-Specific Learning Goals for STEM Integration Using ViMAP-Tangible

The curricular activities consisted of three phases, and were designed to integrate multiple domain-specific learning goals. These learning goals are shown in Figure 1.

During the first phase of our curriculum, students were introduced to agent-based programming using ViMAP. Students learned to generate "open" and "closed" geometric shapes (e.g. squares, circles, spirals) using ViMAP. In the second phase, students used these shapes to represent models of phenomena involving continuous change over time. We watched segments of the movie *The Lorax* in class, after which students identified various events depicted in the movie (such as Lorax running and accelerating, objects in free fall, etc.) that could be modeled using ViMAP shapes. This phase was designed to establish shape drawing as a *consequential* (Gresalfi & Ingram-Noble, 2008) activity that is not only valuable for its visual aesthetics, but that can also be used as a computational model of change over time. In both these phases, each student worked individually. Our previous work has shown that these two phases can effectively introduce students to the basics of agent-based programming, and students also learn to use their programs (i.e., geometric shapes) as models of motion in Phase 2 (Sengupta & Farris, 2012).

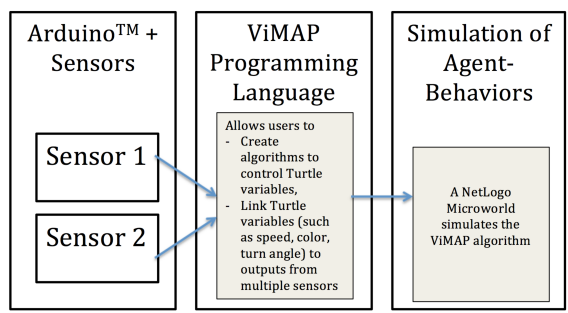

Figure 2: The ViMAP-Tangible Distributed Computation Infrastructure

The third phase used a STEM design challenge (Berland, 2013), in which students worked in dyads to design and develop a tool for shape drawing using ViMAP. The nature of the tool that the dyads were asked to build was a hybrid computational machine consisting of a physical control structure (i.e., a simple machine), and a ViMAP program, which would together control the movement of the virtual agent (ViMAP turtle) on screen. Each student was asked to design a mechanical control for one sensor, but the dyad was responsible for jointly designing the ViMAP program. Emphasis on User Centered Design was maintained throughout this phase as follows: First, students themselves acted as the users; thereafter, students from other groups acted as users (this is the main user-testing phase for the purposes of our analysis); and during the final day of the class, students' parents were invited to visit the class to test their children's designs.

Our pedagogical approach emphasized several of Gulliksen *et al.*'s (2003) core principles of UCD. These included the following: an emphasis on rapid prototyping during the early phases of the design process; a cyclic iterative process of designing solutions interwoven with evaluation; and within group distributed expertise for different design components.

## 4. RESEARCH QUESTIONS

We investigate the following research questions:
1. How are students' computational and mechanical representations shaped by the following curricular foci:
    1.1. User-Centered Design?
    1.2. Collaboration between learners?
2. What is the relationship between the structural characteristics of students' physical and computational inventions and the development of their scientific and mathematical knowledge?

## 5. SETTING & METHOD

The study took place in a metropolitan city in the form of an enrichment program for elementary school children, conducted in a classroom on the campus of a large private university in the mid-southern USA. Classes met once a week (9:00 a.m. to 11:30 a.m.) on six consecutive Saturday mornings. Students were recruited through an online solicitation sent to the local elementary schools. None of the students in this course had any prior programming experience, and we particularly encouraged female students to apply. Students were admitted on a first-come, first-served basis. There were 16 participants, out of which eight students were in 3rd grade, and eight were in 4th grade. Five of the students were female, and three of them were in 3rd grade. The ethnic composition of the students whose work we analyzed for this paper is as follows: White (7), Asian American (4), and African American (3). Two students were absent for multiple days, and their work has not been analyzed.

The first author acted as the lead instructor for the study. Both the authors also collected data in the form of videotaped in-depth interviews with the participants, the software and hardware artifacts (i.e., ViMAP programs and physical machines) designed by the students, and field notes. The interviews were conducted while the learners were engaged in the modeling and programming activities. In some cases, the interviews ensued when the learner called upon the researcher in order to help him or her with a difficulty. In other cases, researchers conducted interviews in order to ask learners to explain their programs or models.

Similar to Carver *et al.* (1992), we believe that the role of the teacher in such a design-based classroom can be best described as using the cognitive apprenticeship framework (Collins, Brown & Newman, 1987). That is, the teacher(s) supported students' development of representational practices in some cases by explicitly *modeling* certain elements of the design practices (e.g., by acting out the ViMAP commands physically in class), and in other cases, prompting students to *reflect* on the changes they are making to their designs in terms of their affordances.

Our data consisted of student-generated artifacts, field notes, and video-recorded interviews of students with the researchers. We analyzed this data by identifying themes and sub-themes using the double coding method (Miles & Huberman, 1994). To present our analysis, we used a case study-based approach. Our selection of cases was guided by the following criteria: *representativeness* and *typicality*. *Representativeness* implies that the selected cases should aptly represent key aspects of the instructional and learning processes. These key aspects, in turn, are defined based on the putative research question(s). *Typicality* implies that the selected case(s) should potentially represent aspects of the process of learning experienced by the majority of the student population.

We answered the research questions (RQs) as follows. In order to investigate the role of UCD and collaboration (RQ1), we compared the intermediate and final products of the students' design activities before and after user testing (during Phase 3) to check for improvements. In order to investigate the affordances of the students' designs in terms of the scientific and mathematical concepts and discourse that the students engaged in their designs (RQ2), we analyze the structure of each group's physical machines, their ViMAP program (i.e., programming commands), and their instructions for users in terms of the type(s) of mathematical measures generated by each group. The children generated these measures in order to provide explicit instructions to users for operating and understanding their machines. These measures in turn were categorized either as a) mathematical i.e., involving either multiplicative or proportionality-based reasoning, or reasoning involved in understanding geometric coordinate systems; or b) physics-based, i.e., indicative of reasoning about the measurement of speed and distance. 20% of the data reported here was blind-coded by an additional coder not involved with our study (Cohen's Kappa = 0.95).

## 6. FINDINGS

### 6.1 Pedagogical Affordances of UCD & Collaboration

Across all the groups, we found that (1) the distributed computing infrastructure resulted in positive interdependence (Kreijns *et al.*, 2003) among students, where members of each group made coordinated, non-redundant contributions towards a common goal; and (2) a focus on User-Centered Design resulted in students refining both the software and hardware components of their programs by engaging in iterative design cycles.

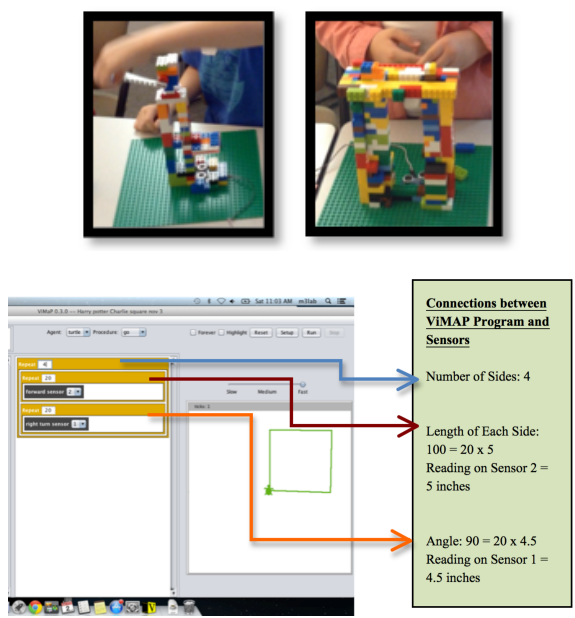

Figure 3A (Top Left): Jerry's pulley mechanism for controlling turn of the turtle via Sensor 1; Figure 3B (Top Right): Chuck's machine for controlling the speed of the turtle via Sensor 2. Figure 3C (Bottom) is a screenshot of their ViMAP program for generating a square, and shows how it responds to the sensors.

An illustrative case is the work of Chuck and Jerry (Figure 3A, 3B & 3C). During Phase 3, they built two separate machines to represent specific hand movements over the sensors. Jerry's machine consisted of a flat LEGO plate that could be lowered or raised above an ultrasonic distance sensor (Sensor 1 in Figure 3), using a pulley mechanism, to control the *turn angle* of the computational agent. Chuck's machine also comprised of a horizontal LEGO surface that could be lowered or raised above a distance sensor (Sensor 2 in Figure 3), using a manually operated crank lift, to control the *step-size* of the turtle.

In order to facilitate better collaboration between group members, during Phase 3, students' ViMAP programs became iteratively more refined with fewer bugs and redundancies. For example, in their initial versions, Chuck and Jerry had linked multiple turtle variables to each of the sensors, and after a few attempts, realized that their coordinated actions would make such a design redundant. Instead, they decided to divide their responsibilities: one person would control the turn, while the other would control the speed of the turtle, thereby, creating *positive interdependence*. Another effect of collaboration was the shared development of mathematical measures within each group, which in turn fostered mathematical and physics-related discourse and representational practices. This is discussed in Section 6.2.

We also found that students' ViMAP programs and instructions for users, post user-testing, were more generative, and communicative. We found that the final designs of six out of seven groups allowed users to draw *multiple* shapes, whereas their initial designs during Phase 3 were more constrained and could only generate a specific shape (typically, a circle). This improvement was a direct result of user testing, as nearly every user demanded to be able to draw more than one shape using the same tool. In some cases we also noticed that the reliability of the output of their ViMAP program also improved as a result of improvement of their physical structures. For example, both Chuck and Jerry realized that they had to improve the flatness of the surfaces that were generating the sensor-readings, because their users were unable to generate reliable outputs. This resulted in Chuck introducing a flat paper strip to cover the bottom of the LEGO plates, while Jerry created a wider plate to control sensor readings more reliably.

## 6.2 Relationship Between Children's Inventions & Learning Math and Physics

### 6.2.1 Designs that primarily supported learning rate and kinematics

Figure 4 shows an example of student work in which two students, Seana and Curly, built two separate, manually operated, wheeled cars, with flat surfaces in front (stacked LEGO bricks) to represent the palm of a hand. The operating mechanism involves pushing the car towards or away from the sensor, where distance of the flat surface (representing the "palm of a hand") from the sensor generates the reading of the ultrasonic sensor. The distance of one of the cars from the sensor controls the speed of the ViMAP turtle, while distance of the other car from the second sensor controls the rotation of the turtle. A total of two groups of students developed two cars as their drawing machines.

After user-testing, Seana and Curly realized that they had to provide instructions to the user on how fast they would have to move each car in order to generate the desired shape(s). This is because, in their design, moving the two cars at different rates resulted in different shapes. Neither student had any experience in calculating rates or speed prior to this study, nor were they formally instructed during this class to calculate the rates. However, they used the computer clock and cell phones as timers, and figured out by trial and error how fast they had to move each car in terms of the time taken by their cars to travel specific distance(s) in order to generate the desired shape(s). This also resulted in making their designs more communicative by annotating the track at specific positions, along with some written and verbal instructions for users regarding how fast they needed to move the cars between these annotated positions. Although this was a rough measure of rate, we believe that this is a productive entry point into learning kinematics through the development of consequential physics talk, as well as mathematical discourse on rates – i.e., through engaging in "conversations for conceptual change" (diSessa *et al*., 1991).

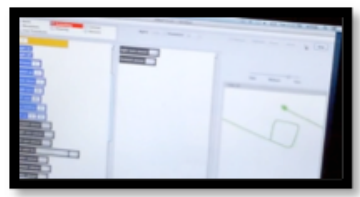

Figure 4A: Seana and Curly's ViMAP program for shape drawing

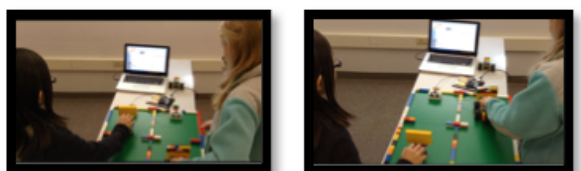

Figure 4B (Left) and 4C (Right): Seana (dressed in black) and Curly (dressed in blue and white) take turns in moving their "cars" towards the ultrasonic sensors. Seanna controls the rotation, and Curly controls the step-size of the Turtle.

### 6.2.2 Designs that primarily supported multiplicative and proportionality-based reasoning

An example of this type of design is Chuck and Jerry's work described in Section 6.1. Jerry's machine controlled the angular turn, and the user could generate readings from 1 – 10 (inches) by using the pulley mechanism. However, the user could also alter the numerical parameter of the "Repeat" command in the ViMAP program, which in turn would effectively multiply the sensor reading by that parameter. Chuck used a similar strategy to let users control the step-size of the turtle; he had created several visible marks on his towers at increments of a third of the maximum height, so that the user can generate shapes of three levels of magnification.

Chuck and Jerry, who were both beginning to learn multiplication tables in their regular math class, thus got an opportunity to use and further develop their multiplicative and proportionality-based reasoning in order to make their designs work. We found that a total of three groups of students also invented this type of design.

### 6.2.3 Designs that primarily supported learning the Cartesian coordinate system

Two groups of students developed machines similar to Chuck and Jerry, but using a Cartesian coordinate system in their ViMAP program. One of their machines controlled the translational displacement of the turtle, while the other controlled the vertical displacement of the turtle. To do so, they used ViMAP commands such as "jump-X-by <Sensor-reading>" and "jump-Y-by <Sensor-reading>", respectively. An illustrative example is the case of Ken & Yang. As they iteratively refined the various combinations of translational and vertical displacements that would generate the different shapes, they became familiarized with the NetLogo XY-coordinate space. They realized during user-testing that it is much easier for users to control their machines if they are provided with coordinates for the vertices of target shapes such as equilateral triangles and rectangles. This was evident in their instructions for users on the final day.

## 7. CONCLUSION & DISCUSSION

In this paper, we have presented a socio-technically distributed pedagogy for Integrated STEM in elementary grades. Integrated STEM is a necessarily interdisciplinary enterprise, and given the nature of elementary classrooms (K-4), where the same teacher is responsible of teaching all disciplines, we believe that K-4 is a great setting for such curricula. The cases we presented show that 3$^{rd}$ and 4$^{th}$ grade students were indeed able to successfully participate productively in an engineering design process that highlights User Centered Design and collaboration – key elements of the practice of engineering. Furthermore, we also showed that in the process, students were able to iteratively develop expertise in agent-based programming, as well as mathematical and scientific discourse and representational practices such as measurement and modeling.

Our paper also represents the first attempt to integrate agent-based computational thinking with engineering, curricular science and math within a single curriculum in elementary grades. To this end, we presented a pedagogical innovation in the form of new socio-technically distributed computational environment - ViMAP-Tangible - and a set of curricular activities in the genre of STEM Design Challenges (Berland, 2013). The computational abstractions in the form of ViMAP programming language, as well as the associated epistemic and representational practices such as measurement and modeling - act as *invariants* that facilitate *cohesion* across the STEM domains (Nathan *et al.*, 2013).

Our curricular design proposes a mechanism for bringing about such cohesion using a socio-technically distributed computational infrastructure. Following Berland (2013), we maintained focus on engineering practices from the beginning (Phases 1 & 2) by engaging students in design, in the form of iterative development and refinement of their computational models. Furthermore, we leveraged the synergy between tangibility and collaborative learning in a manner that introduced students to another key engineering practice (i.e., collaborative design). This in turn supported the development of scientific and mathematical knowledge by creating opportunities for reflection and positive interdependence.

## ACKNOWLEDGEMENTS

Financial support was provided by the National Science Foundation (NSF Early CAREER # 1150230).